\documentclass[twocolumn,showpacs,aps,prl,superscriptaddress]{revtex4}

\usepackage{graphicx}
\usepackage{dcolumn}
\usepackage{amsmath}
\usepackage{units}

\input{babarsym}

\newcommand{\microns}{\ensuremath{\mu{\rm m}}}
\newcommand{\Btag}{\ensuremath{\B_\mathrm{tag}}}

\newcommand{\Bztokspiz} {\ensuremath{\Bz \to \KS\piz}}
\newcommand{\Bztoksksks} {\ensuremath{\Bz \to \KS\KS\KS}}
\newcommand{\cksksks} {\ensuremath{C}}
\newcommand{\sksksks} {\ensuremath{S}}

\newcommand{\Bflav} {\ensuremath{B_{\rm flav}}}
\newcommand{\zrec}{\ensuremath{z_{\CP}}}
\newcommand{\ztag}{\ensuremath{z_\mathrm{tag}}}

\newcommand{\BABARPubYear}    {04}
\newcommand{\BABARPubNumber}  {xxx}
\newcommand{\SLACPubNumber} {xxxxx}

\def\figurebox#1#2#3{%
    \def\arg{#3}%
    \ifx\arg\empty
    {\hfill\vbox{\hsize#2\hrule\hbox to #2{\vrule\hfill\vbox to #1{\hsize#2\vfill}\vrule}\hrule}\hfill}%
    \else
    {\hfill\epsfbox{#3}\hfill}%
    \fi}

\long\def\inst#1{\par\nobreak\kern 4pt\nobreak
    {\it #1}\par\vskip 10pt plus 3pt minus 3pt}

\begin{document}

\preprint{\babar-PUB-\BABARPubYear/\BABARPubNumber}
\preprint{SLAC-PUB-\SLACPubNumber}

\title{
  { \Large \bf \boldmath Branching Fraction and \CP\ Asymmetries of \Bztoksksks  }
}
%% author list as of 02-Dec-2004 (617 authors)
%
\author{B.~Aubert}
\author{R.~Barate}
\author{D.~Boutigny}
\author{F.~Couderc}
\author{Y.~Karyotakis}
\author{J.~P.~Lees}
\author{V.~Poireau}
\author{V.~Tisserand}
\author{A.~Zghiche}
\affiliation{Laboratoire de Physique des Particules, F-74941 Annecy-le-Vieux, France }
\author{E.~Grauges-Pous}
\affiliation{IFAE, Universitat Autonoma de Barcelona, E-08193 Bellaterra, Barcelona, Spain }
% ...
\collaboration{The \babar\ Collaboration}
\noaffiliation

%\collaboration{The \babar\ Collaboration}

\date{\today}

\begin{abstract}
  We present measurements of the branching fraction and 
  time-dependent \CP-violating 
  asymmetries in \Bztoksksks\ decays based on 227 million $\Y4S\to\BB$
  decays collected with the \babar\ detector at the PEP-II
  asymmetric-energy $B$ Factory at SLAC. 
  We obtain a branching fraction of
  $(6.9^{+0.9}_{-0.8}\pm 0.6)\times 10^{-6},$ 
  and \CP\ asymmetries 
  $\cksksks = -0.34^{+0.28}_{-0.25} \pm 0.05$ and 
  $\sksksks = -0.71^{+0.38}_{-0.32} \pm 0.04,$ where the first
  uncertainties are statistical and the second systematic.
\end{abstract}


\maketitle

The amplitude of time-dependent \CP violation (CPV) predicted for
$b \to c\cbar\s$ decays of neutral \B mesons in the Standard Model (SM) 
is $\sin2\beta$ where $\beta = {\rm arg}(-V_{cd}V_{cb}^{\ast}/V_{td}V_{tb}^{\ast})$
is the \CP violating phase difference between mixing and
decay amplitudes, with $V_{ij}$ the elements of the Cabibbo-Kobayashi-Maskawa (CKM)
quark mixing matrix~\cite{CKM}.
This prediction has been well tested
at the \B factories in recent years~\cite{BaBarSin2betaObs}.
The SM also predicts the amplitude of CPV in 
$b \to s\qbar\q$ $(\q=d,s)$ decays,
defined as $\sin2\beta_{\rm eff},$
to be approximately $\sin2\beta .$
However, since $b \to s\qbar\q$ decays are dominated by one-loop transitions that can
potentially accommodate large virtual particle masses, contributions from
physics beyond the SM could invalidate this prediction, making these
decays especially sensitive to new physics~\cite{Grossman:1996ke}. 
An active program has arisen to measure $\beta_{\rm eff}$ in as
many $b \to s\qbar\q$ ``penguin'' modes
as possible~\cite{hoecker_ichep}. 
However, many of these final states are affected by additional
SM physics contributions that obscure the measurement of $\beta_{\rm eff}$~\cite{ref:glnq},
or are not \CP eigenstates. 
Two decays to \CP eigenstates that have been noted as having small theoretical
uncertainties in the measurement of $\beta_{\rm eff}$ are
$\Bz \to \phi K^0_s$~\cite{Abe:2003yt,ref:stefanprl,ref:cc} (\CP-odd)
and \Bztoksksks\ (\CP-even)~\cite{ref:gershon}.

In this Letter we present a measurement of time-dependent \CP-violating
asymmetries in the decay \Bztoksksks, along with a measurement of
the branching fraction (BF). 
Until recently the small branching fraction~\cite{ref:3ksbf} and the 
absence of charged decay tracks originating at the \Bz decay vertex
have limited the ability to extract \CP parameters from 
\Bztoksksks. 
However, techniques recently developed
to deal with the reconstruction of the 
\Bz decay vertex in \Bztokspiz\  
have made this measurement 
possible~\cite{ref:k0spi0prl}. 

The time-dependent \CP\ asymmetry is obtained by measuring 
the proper-time difference $\deltat\equiv t_{\CP}-t_\text{tag}$ between a 
fully reconstructed decay \Bztoksksks{}  and the partially reconstructed tagging
\B{} meson (\Btag ).
The asymmetry in the decay rate $f_+$ $(f_-)$ when the tagging meson is a 
\Bz{} (\Bzb ) is given as
\begin{eqnarray}
  \label{eqn:td}
  \lefteqn{ f_{\pm}(\deltat) \; = \; \frac{e^{-|\deltat|/\tau}}{4\tau} \times }\; \\
   && \left[ \: 1 \; \pm \;
     \: S \sin{( \deltamd\deltat)} \mp C \cos{( \deltamd\deltat)} \: \: \right] \; , \nonumber
\end{eqnarray} 
\noindent where the parameters $C$ and $S$ describe the amount of
\CP violation in decay and in the interference between decay with and without mixing,
respectively.
Neglecting CKM-suppressed amplitudes, we expect
$\sksksks=-\sin2\beta$ and $\cksksks=0$ in the SM.

The results presented here are based on $226.6\pm 2.5$ million $\Y4S\to\BB$
decays collected 
with the \babar\ detector at the
PEP-II asymmetric-energy $\epem$ collider, located at the Stanford Linear Accelerator Center. 
The \babar\ detector~\cite{ref:babar} provides charged-particle tracking through a
combination of a five-layer double-sided silicon microstrip
detector (SVT) and a 40-layer central drift chamber, both
operating in a \unit[1.5]{T} magnetic field
Charged kaon and pion identification is achieved through measurements of particle energy-loss in
the tracking system and Cherenkov cone angle 
in a detector of internally reflected Cherenkov light.  A
segmented CsI(Tl) electromagnetic calorimeter provides
photon detection and electron identification.  Finally, the
instrumented flux return of the magnet allows discrimination
of muons from pions.

Candidates for \Bztoksksks{} are formed by
combining three \KS{} candidates in an event. 
We reconstruct
$\KS\to\pip\pim$ candidates from pairs of oppositely charged
tracks. The two-track combinations must form a vertex with a 
$\pip\pim$ invariant mass within $12\mevcc$ 
(about $4\sigma$) 
of the nominal \KS\ mass~\cite{Hagiwara:fs}, 
a reconstructed flight
distance between 0.2 and 40.0\cm\ from the beam spot in the plane
transverse to the beam, 
and an angle between the
transverse flight direction and the transverse momentum vector of
less than 200\mrad. 
For each $B$
candidate two nearly independent kinematic variables are computed,
namely the beam-energy-substituted mass 
$\mes=\sqrt{(s/2+{\bf p}_i \cdot {\bf p}_B)^2/E_i^2+p^2_B}$, 
and the energy difference $\DeltaE=E^*_B-\sqrt{s}/2$. 
Here, $(E_i,{\bf p}_i)$ is the
four-vector of the initial \epem{} system, $\sqrt{s} $
is the center-of-mass energy, ${\bf p}_B$ is the
reconstructed momentum of the \Bz{} candidate, and $E_B^*$ is its
energy calculated in the \epem{} rest frame. For signal decays,
the \mes{} distribution peaks near the \Bz{} mass with an rms
deviation of about \unit[$2.5$]{\mevcc} and the \DeltaE{}
distribution peaks near zero with an rms deviation of about \unit{$14$}{\mev}. 
We select candidates within the window
\unit[$5.22<\mes<5.30$]{\gevcc} and
\unit[$-120<\DeltaE<120$]{\mev}, which includes the signal peak
and a ``sideband'' region for background characterization. 

The sample of \Bztoksksks{} candidates is dominated by random
$\KS\KS\KS$ combinations from $\epem\to\qqbar$ $(\q=u,d,s,c)$
fragmentation. Monte Carlo (MC) studies show that contributions from
other \B{} meson decays can be neglected. We exploit topological
observables to discriminate the jet-like $\epem\to\qqbar$ events
from the more uniformly distributed \BB{} events. In the $\Y4S$
rest frame we compute the angle $\theta^*_T$ between the thrust
axis of the \Bz{} candidate and that of the remaining particles in
the event.  While $|\cos\theta^*_T|$ is highly peaked near 1 for
$\epem\to\qqbar$ events, it is nearly uniformly distributed for
\BB{} events.  We require $|\cos\theta^*_T|<0.9$, eliminating $\sim 68\%$ of
the background.  
In addition we use a Fisher discriminant variable $({\cal F})$, 
based on the momenta and angles of tracks in the event~\cite{ref:k0spi0prl}, 
in the maximum-likelihood fit described below.

For the~\unit[$1.4$]{\%} of events with more than one candidate we
select the combination with the smallest $\chi^2=\sum_{i}
(m_i-m_{\KS})^2/\sigma^2_{m_i}$, where $m_i$ ($m_{\KS}$) is the measured
(nominal $\KS$) mass and $\sigma_{m_i}$ is the estimated uncertainty on
the mass of the $i$th $\KS$ candidate. 
We also remove all \Bz{} candidates
that have a $\KS\KS$ mass combination within $3\sigma$
$(45\mevcc)$ of the $\chi_{c0}$ or $\chi_{c2}$ mass. While we
expect few $\chi_{c0}$ and $\chi_{c2}\to\KS\KS$ decays in our final
sample, these are $b \to c\cbar\s$ decays that would bias 
the \CP-asymmetry measurement.

We extract the results from unbinned maximum-likelihood 
fits to the kinematic, event shape $({\cal F})$, and $\deltat$ variables.
We maximize the logarithm of an extended likelihood function
\begin{eqnarray}
{{\cal L} = 
%%%\frac{
e^{-(N_S+N_B)}
%%%}{(N_T)!}  
\times}  
{\prod_i^{N_T}{
}
  \left[ N_S 
{\cal P}^i_{S} +
    N_B 
{\cal P}^i_{B} \right]  ,}  
\nonumber
\end{eqnarray}
\noindent where 
${\cal P}_S$ and ${\cal P}_B$ are
the probability density functions (PDFs) for signal ($S$) and 
continuum background ($B$), $N_T$ is the total number of events,  
and $N_S$ and $N_B$ are the event yields to be
determined from the fit.
The product is over the selected events.
The observables are sufficiently uncorrelated
that we can construct the likelihoods as the products of one-dimensional PDFs.
The PDFs for signal are parameterized from signal MC events.  
For background PDFs we
determine the functional form from data in the sideband regions of the
other observables where backgrounds dominate.  
We include these regions in the fitted sample 
and simultaneously extract the parameters
of the background PDFs along with the fit results.

\begin{figure}[!tbp]
\begin{center}
\includegraphics[width=1.0\linewidth]{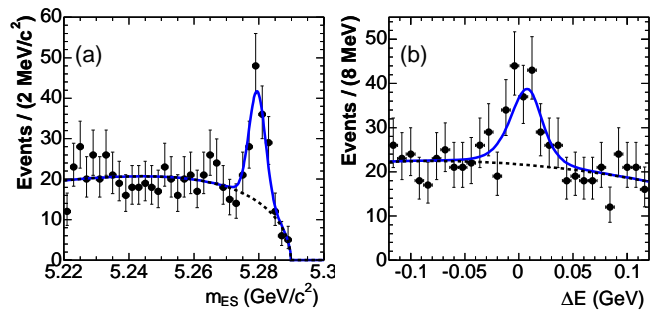}
\caption{Distribution of (a) $\mes$ and (b) $\DeltaE$ for all events
that pass the selections used for determining the branching fraction. 
The solid (dashed) curves are the PDF projections for the signal 
plus background (background only) from the fit.} 
\label{fig:prplots}
\end{center}
\end{figure}

For the branching fraction fit 
we use only the kinematic and event-shape variables 
$({\cal P}_{BF} = {\cal P}(\mes ) {\cal P}(\DeltaE ) 
{\cal P}({\cal F})).$
There are two yields and six
continuum PDF parameters floated in the fit.
There are 721 $\KS\KS\KS$ candidates that pass all the above criteria, 
and the fit to this data yields $N_S=88\pm 10$ events and $N_B=633\pm 26$ events.
Figure~\ref{fig:prplots} shows the \mes{} and $\DeltaE$ distributions for these
events with the results of the fit plotted as curves.
As a check we also add a fit component for random combinatorial \B background, 
with PDF parameters determined from large MC samples. 
This fit finds $14\pm 11$ candidates assigned to the \B\ background.
These candidates come from the continuum background; the signal yield
changes by less than one candidate.
A signal reconstruction efficiency of $5.6\%$ is derived from a large MC sample
in which the $\KS$ reconstruction efficiency is carefully matched with that observed 
in large hadronic data samples. 
Assuming equal production rates of $B^0\overline{B}^0$ and $B^+B^-,$ 
we determine ${\cal B}(\Bztoksksks )=(6.9^{+0.9}_{-0.8}\pm 0.6)\times 10^{-6}.$

The largest systematic error $(5\% )$
for the branching fraction measurement
comes from our uncertainty on the efficiency of
reconstructing $\KS\to\pi^+\pi^-$ decays. 
We determine uncertainties of $4\% $ for the 
effect of the candidate selection cuts
and $5\% $ for the parametrization of the PDFs used in the fit.
The remaining uncertainties, 
including possible error in modeling 
the $\KS\KS\KS$ Dalitz plot distribution in determining the signal efficiency, 
combine to $2\% .$

The \CP-fit PDF for a given tagging category is 
${\cal P}_{\CP}^c={\cal P}_{BF}{\cal P}^c(\deltat,\sigma_{\deltat})\epsilon^c$
where $\epsilon^c$ is the tagging efficiency for tag category $\c$.  
The total likelihood $\cal L$ is the product of
likelihoods for each tagging category, 
and the free parameters are determined by maximizing the quantity $\ln \cal L$.  
Along with the CPV asymmetries $S$ and $C$, 
the fit extracts $\epsilon^c$ for the background and other background parameters.  
The background PDFs include parameters for the \deltat-resolution function ${\cal R}$ 
and for asymmetries in the rate of \Bz{} versus \Bzb{} tags.  
We extract 25 parameters from the \CP fit.

We use a neural network to determine the flavor of the $\Btag$
meson from kinematic and particle-identification
information~\cite{ref:sin2betaPRL02}. 
Each event is assigned to one of
six mutually exclusive tagging categories, designed to combine flavor
tags with similar performance and \deltat\ resolution.  We
parameterize the performance of this algorithm with a data sample
($B_{\rm flav}$) of fully reconstructed $\Bz\to D^{(*)-}
\pip/\rho^+/a_1^+$ decays. The effective tagging efficiency
obtained from this sample is $Q\equiv\sum_c \epsilon^c
(1-2w^c)^2=0.305\pm 0.004$, where $\epsilon^c$ and $w^c$ are the
efficiencies and mistag probabilities, respectively, for events tagged
in category $c$. 

We compute the proper-time difference $\deltat=(\zrec-\ztag)/\gamma\beta c$
using the known
boost of the \epem{} system and the measured
$\deltaz=\zrec-\ztag$, the difference of the reconstructed decay
vertex positions of the \Bztoksksks{} and \Btag{} candidate along
the boost direction ($z$).  
A description of the inclusive reconstruction of the \Btag{} vertex 
is given in Ref.~\cite{ref:Sin2betaPRD}.  
For the \Bztoksksks{} decay, where no
charged particles are present at the decay vertex, we constrain the
\B meson production vertex to the interaction point (IP) in the transverse
plane using a geometric fit.  The position and size of the interaction region 
are determined on a run-by-run basis from the spatial distribution of vertices from
two-track events.  The uncertainty on the IP position, which
follows from the size of the interaction region, is about
\unit[150]{$\mu$m} horizontally and \unit[4]{$\mu$m} vertically.  
The uncertainty on \zrec, a convolution of the interaction 
region and the vertex of the \Bztoksksks{} decay, is about 75{\microns}.
The uncertainty on \ztag\ is about 200{\microns} and thus the 
uncertainty in $\deltaz$ is dominated by the uncertainty in the 
vertex of the tagging decay. The resulting resolution is comparable to that in \Bz\to\jpsi\KS~\cite{ref:k0spi0prl}.

Simulation studies show that the procedure we use to determine the vertex for a
\Bztoksksks{} decay
provides an unbiased estimate of \zrec{}. The estimate of the
$\deltat$ error in an event reflects the strong dependence of the \zrec{}
resolution on the number of SVT
layers traversed by the \KS\ decay daughters. However, essentially all
events have at least one \KS\ candidate for which both tracks have at least
one hit in the inner three SVT layers (at radii from
\unit[$3.2$]{cm} to \unit[$5.4$]{cm}). In this case the mean
\deltat\ resolution is comparable to that in decays in which the
vertex is directly reconstructed from charged particles
originating at the $B$ decay point~\cite{ref:Sin2betaPRD}.  For a
small fraction (0.1\%) of the signal events, at least one \KS\ has tracks
with hits in the outer two SVT layers (at radii \unit[$9.1$]{cm}
to \unit[$14.4$]{cm}) but none of the three \KS\ s have hits in the inner three layers. In this case 
the resolution is nearly two times worse but the event can still be used in the \CP fit. 
Events with \unit[$\sigma_{\deltat}>2.5$]{ps} or \unit[$|\deltat|>20$]{ps}
are excluded from the \CP\ fit.

The resolution function ${\cal R}$ is parameterized as the sum of a `core' and a
`tail' Gaussian distribution, each with a width and mean proportional to 
$\sigma_{\deltat}$, and a third Gaussian with a mean of
zero and a width fixed at \unit[$8$]{ps}~\cite{ref:Sin2betaPRD}.  
We have verified with MC simulation that the parameters of ${\cal R}
$ for \Bztoksksks\ decays are similar to those obtained from the $B_{\rm flav}$ sample.
Therefore, we extract these parameters from a fit to the $B_{\rm flav}$ sample.  
We find that the \deltat{} distribution of background candidates is well described by a
delta function convolved with a resolution function having the same
functional form as that for the signal. The parameters of the
background function are determined in the fit.

\begin{figure}[!tbp]
\begin{center}
\parbox{0.47\textwidth}{\includegraphics[width=0.9\linewidth]{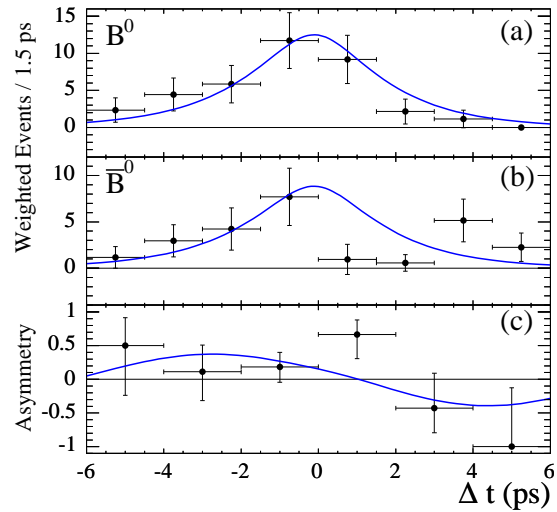}}
\end{center}
\caption{
  Distributions of $\deltat$ for events weighted with the sPlot technique for
  $B_{\rm tag}$ tagged as (a) $\Bz$ or (b) $\Bzb$, and (c) the
  asymmetry ${\cal A}(\deltat)$.  The points are weighted data
  and the curves are the PDF projections. }
\label{fig:dtplot}
\end{figure}

The fit including $\deltat$ and tagging information yields
\mbox{$\sksksks =
  -0.71^{+0.38}_{-0.32}\pm 0.04$} and \mbox{$\cksksks =
  -0.34^{+0.28}_{-0.25}\pm 0.05$}.
Fixing \mbox{$\cksksks=0$} we obtain
\mbox{$\sin 2\beta=-\sksksks=0.79^{+0.29}_{-0.36}\pm 0.04$}. 
Figure~\ref{fig:dtplot} shows distributions of
$\deltat$ for $\Bz$-tagged and $\Bzb$-tagged events, and the asymmetry
${\cal
  A}(\deltat) = \left( N_{\Bz} -
  N_{\Bzb}\right) /\left( N_{\Bz} + N_{\Bzb}\right) ,$ 
obtained with the sPlot event weighting technique~\cite{splot}.

\begin{table}[htb]
  \caption{Systematic uncertainties on $S$ and $C$.}
  \centerline{\small
    \begin{tabular}{lcccc}
      \hline\hline
      {}              &  \multicolumn{2}{c}{$\sigma (S)$}  & \multicolumn{2}{c}{$\sigma (C)$} \\
      \hline
      Resolution function      & \multicolumn{2}{c}{0.017}  & \multicolumn{2}{c}{0.017} \\
      Vertex reconstruction         & \multicolumn{2}{c}{0.020}  & \multicolumn{2}{c}{0.022} \\
      SVT alignment            & \multicolumn{2}{c}{0.015}  & \multicolumn{2}{c}{0.008} \\
      Background asymmetry     & \multicolumn{2}{c}{0.007}  & \multicolumn{2}{c}{0.022} \\
      Fit correlation        & \multicolumn{2}{c}{0.016}  & \multicolumn{2}{c}{0.004} \\
      Tag-side interference    & \multicolumn{2}{c}{0.008}  & \multicolumn{2}{c}{0.015} \\
      PDFs                     & \multicolumn{2}{c}{0.025}  & \multicolumn{2}{c}{0.026} \\
      \hline
      Total             &\multicolumn{2}{c}{0.044} &\multicolumn{2}{c}{0.047} \\
      \hline\hline
    \end{tabular}}
  \label{tab:summsys}
\end{table}

Systematic uncertainties on the \CP\ parameters are given in Table I.
The systematic errors are evaluated with large samples
of simulated \Bflav\ and \Bztoksksks\ decays. 
We employ the
difference in resolution function parameters extracted from these
samples to vary the resolution function
parameters extracted from the $B_{\rm flav}$ sample in data. 
We also perform fits
to the simulated \Bztoksksks\ signal with parameters
obtained either from signal or \Bflav\ events to account for any
potential bias due to the vertexing technique. 
Several SVT misalignment scenarios are applied to the simulated \Bztoksksks\ events to estimate
detector effects.
We consider large variations of the IP position and resolution and
find they have negligible impact. 
Asymmetries in the rate of \Bz{} versus \Bzb{} tags in the background events, 
which are free parameters
in the fit, are fixed to zero as a systematic uncertainty.
The systematic error due to correlations in the fit variables is
extracted from a fit to a sample of randomly selected signal MC events 
added to background events from a parametrized MC.
We allow for the
possible interference between the suppressed $\bbar \to \ubar c \dbar$ 
and the favored $b \to c \ubar d$ amplitude
for some tag-side \B decays~\cite{ref:tagint}.
Finally, we include a systematic uncertainty to account for imperfect knowledge of
the PDFs used in the fit.  
Most of the uncertainties on the PDFs are statistical and some are associated
with data and MC differences.  

In summary, we have measured the \Bztoksksks\ branching fraction 
and the time-dependent CPV asymmetries.
The BF measurement is in good agreement with previous measurements~\cite{ref:3ksbf}.
The measurements of $\sksksks$ and $\cksksks$ are in good agreement with the SM expectation.

\par
We are grateful for the excellent luminosity and machine conditions
provided by our \pep2\ colleagues, 
and for the substantial dedicated effort from
the computing organizations that support \babar.
The collaborating institutions wish to thank 
SLAC for its support and kind hospitality. 
This work is supported by
DOE
and NSF (USA),
NSERC (Canada),
IHEP (China),
CEA and
CNRS-IN2P3
(France),
BMBF and DFG
(Germany),
INFN (Italy),
FOM (The Netherlands),
NFR (Norway),
MIST (Russia), and
PPARC (United Kingdom). 
Individuals have received support from CONACyT (Mexico), A.~P.~Sloan Foundation, 
Research Corporation,
and Alexander von Humboldt Foundation.


\begin{thebibliography}{99}
\bibitem{ref:cc} Unless explicitly stated otherwise, conjugate decay modes are
assumed throughout this paper.
\end{thebibliography}
\end{document}